\documentclass[prb,aps,amssymb,shownopacs,superscriptaddress]{revtex4}

\usepackage{amsmath}
\usepackage{amssymb}
\usepackage{amsthm}
\usepackage{amsfonts}
\usepackage{listings}
\usepackage{bbold}
\usepackage{enumerate}
\usepackage{latexsym}

\usepackage{psfrag}

\usepackage{bm}
\usepackage[pdftex]{graphicx}
\usepackage{subfigure}

\usepackage{hyperref}

\usepackage{amsmath,amscd}

\def\l{\left}
\def\r{\right}

\def\bs{\boldsymbol}

\DeclareMathOperator{\ptn}{Par}

\newcommand{\mC}{\mathcal{C}}
\newcommand{\mD}{\mathcal{D}}

\newcommand{\prf}[1]{\noindent\begin{proof}#1\end{proof}}

\newcommand{\arr}[2]{\left\{\hspace*{-0.15cm}\begin{array}{#1}#2\end{array}\right.}

\newcommand{\eqn}[2]{\begin{align}\begin{split}\label{#2}#1\end{split}\end{align}}

\newcommand{\eq}[2]{\begin{align*}\begin{split}\label{#2}#1\end{split}\end{align*}}

\theoremstyle{theorem}
\newtheorem{theorem}{Theorem}
\newcommand{\thm}[2]{\begin{theorem}\label{#2}#1\end{theorem}}

\theoremstyle{lemma}
\newtheorem{lemma}{Lemma}
\newcommand{\lma}[2]{\begin{lemma}\label{#2}#1\end{lemma}}

\theoremstyle{corollary}
\newtheorem{corollary}{Corollary}
\newcommand{\cor}[2]{\begin{corollary}\label{#2}#1\end{corollary}}

\newtheorem{conjecture}{Conjecture}
\newcommand{\con}[2]{\begin{conjecture}\label{#2}#1\end{conjecture}}

\theoremstyle{definition}
\newtheorem{definition}{Definition}

\theoremstyle{definition}
\newtheorem{example}{Example}
\newcommand{\ex}[2]{\begin{example}\label{#2}#1\end{example}}

\theoremstyle{theorem}
\newtheorem{notice}{Note}

\begin{document}

\tolerance 10000

\newcommand{\vk}{{\bf k}}

\title{On the particle entanglement spectrum of the Laughlin states}

\author{B.~Majidzadeh Garjani}
\affiliation{Department of Physics, Stockholm University, SE-106 91 Stockholm, Sweden}
\affiliation{%
Nordita, Royal Institute of Technology and Stockholm University, Roslagstullsbacken 23, SE-10691 Stockholm, Sweden
}
\author{B.~Estienne}
\affiliation{Sorbonne Universit\'es, UPMC Univ Paris 06, UMR 7589, LPTHE, F-75005, Paris, France}
\affiliation{CNRS, UMR 7589, LPTHE, F-75005, Paris, France}
\author{E.~Ardonne}
\affiliation{Department of Physics, Stockholm University, SE-106 91 Stockholm, Sweden}

\begin{abstract}
The study of the entanglement entropy and entanglement spectrum has proven to
be very fruitful in identifying topological phases of matter. Typically, one performs
numerical studies of finite-size systems. However, there are few rigorous results in this regard. We revisit the problem of determining the rank of the
``particle entanglement spectrum" of the Laughlin states. We reformulate the
problem into a problem concerning the ideal of symmetric polynomials that vanish
under the formation of several clusters of particles. We introduce an explicit generating set of this ideal, and we prove that 
polynomials in this ideal have a total degree that is bounded from below. We discuss the difficulty in proving the same bound on the degree of any of the variables, which is necessary to determine
the rank of the particle entanglement spectrum.  
 \end{abstract}

\date{\today}

\maketitle

\section{Introduction}

The study of topological phases of matter has benefited greatly from considering the
entanglement properties of the ground states of topological phases. The work of
Kitaev and Preskill\cite{kitaev2006} and of Levin and Wen\cite{lw06} revealed that
the entanglement entropy is a good probe of the topological nature of a system and provides a measure for the
particle content of the topological phase\cite{k06}.The entanglement entropy of a pure quantum state $| \Psi \rangle$ relative to a bipartite partition of the total Hilbert space $\mathcal{H} = \mathcal{H}_A \otimes \mathcal{H}_B$ provides a measure of the entanglement of $| \Psi \rangle$. The entanglement entropy is defined as the Von Neumann entropy $S$ of the reduced density matrix of either one of the two parts. For instance
\begin{align}
S =  - \text{Tr} (\rho_A \log \rho_A), 
\end{align}
where $\rho_A =   \text{Tr}_B \left( | \Psi \rangle \langle \Psi | \right)$.

 In the context of the fractional quantum Hall (FQH) effect, various ways to partition the Hilbert space were proposed\cite{zozulya07}.
Of particular importance, is the spatial partitioning scheme in which the system is split into two regions $A$ and $B$ separated
by a real-space cut of length $L$. For a system exhibiting topological order the real-space entanglement entropy is of the
form\cite{k06,lw06}
\begin{align}
S = \alpha L - \gamma + \cdots, \label{area law}
\end{align}
where $\cdots$ stands for subdominant terms as $L$ becomes large.  The subdominant term $\gamma$ is universal, and depends only on the nature of the topological phase. It bears the name  \emph{topological entanglement entropy}, and is a measure for the particle content of the topological phase\cite{k06}.  The first term $\alpha L$, while non-universal, means that the amount of entanglement is proportional to the length of the boundary separating the two regions. This property called \emph{area law} has appeared in various areas of physics, such as black-hole physics and quantum information. For a quantum many-body state, this property is of particular importance since it opens the way to extremely efficient numerical simulations such as the
Density Matrix Renomalization Group\cite{white92} and Matrix Product States\cite{perez07} methods.
For FQH state this avenue of research was successfully undertaken\cite{zaletel12,estienne13,zaletel13}
and opened the way to a reliable microscopic calculation of quasi-holes properties such as radius and
braiding\cite{wu-14-unpublished}.

Although the real-space cut is of paramount importance in the study of topological phases of matter,
there are other natural ways to partition a quantum Hall system: the \emph{orbital cut}, and the \emph{particle cut}\cite{zozulya07}.
While, in principle, the entanglement entropy behaves according to the
area law
Eq.~\eqref{area law} only for real-space cuts,
it was numerically observed\cite{haque07} that the area law is also valid for orbital cuts. In this paper we will concentrate
on the particle cut, in which one numbers the (identical) particles constituting the phase (for instance, the electrons in the
quantum Hall case), and one declares the particles numbered $1,2,\ldots,N_A$ to belong to subsystem $A$, while the
remaining particles numbered $N_A+1$, $N_A+2$, \ldots, $N$ belong to subsystem $B$.
The spectrum of the reduced density matrix obtained by tracing out the particles in subsystem $B$ is the
``particle entanglement spectrum" (PES)\cite{sterdyniak11}. 

While the entanglement entropy $S$ provides a good probe of topological order, the topological entanglement entropy $\gamma$
does not determine unambiguously the universality class of the topological state.
Li and Haldane\cite{li08} realized that the spectrum of \mbox{$H_A = - \log \rho_A$} itself contains much more information
than the entanglement entropy. They proposed to use the low lying part of this \emph{entanglement spectrum} as a
``fingerprint'' of the topological phase.
To be more specific, under a bipartition $\mathcal{H} = \mathcal{H}_A \otimes \mathcal{H}_B$, a pure quantum state
$| \Psi \rangle$ admits a Schmidt decomposition
\begin{equation}
| \Psi \rangle = \sum_{i} e^{-\xi_i/2}  |  \psi^A_i\rangle  \otimes | \psi^B_i \rangle,
\end{equation}
where the $e^{-\xi_i/2}$'s are positive numbers called the Schmidt singular values, while
$|\psi^A_i\rangle$ and $ | \psi^B_i\rangle$ form orthonormal sets in
$\mathcal{H}_A$ and $\mathcal{H}_B$, respectively. The reduced density matrix is then simply 
\begin{align}
\rho_A =  \sum_{i} e^{-\xi_i}  |\psi^A_i\rangle  \langle \psi^A_i | = e^{-H_A}.
\end{align} 
The entanglement spectrum is the set of all \emph{entanglement energies} $\xi_i$. The bipartition can be chosen to preserve as much symmetry as possible, which in turn yields quantum numbers for the $\xi_i$'s, such as the momentum along the cut. Li and Haldane observed that\textendash per momentum sector\textendash the number of entanglement energies reproduces exactly the number of gapless edge modes. They proposed that tracing out the degrees of freedom of part $B$ introduces a virtual edge for part $A$. The Li-Haldane conjecture is therefore two-fold. For a FQH state in the thermodynamic limit:
\begin{itemize}
\item[(i)] the entanglement energies and edge modes have the same counting,
\item[(ii)] the entanglement spectrum is proportional to the (edge) CFT spectrum.
\end{itemize}
It is now understood that  (ii) can only hold in the case of a real-space cut, which maintains locality along the
cut\cite{dubail12,sterdyniak12,rodriguez12}. 
For an orbital cut the entanglement Hamiltonian $H_A$ has no reason to be local.
On the other hand the point (i) holds irrespective of the cut for model wave functions that can be written as correlation functions in a CFT.
Such wave functions are precisely of MPS form\cite{dubail12a},
and the CFT Hilbert space provides a one-to-one mapping\cite{estienne12} 
between edge modes and entanglement energies.

While the agreement between the counting of the number of modes in the entanglement spectrum and the counting of the edge excitations is well understood, in practice, this fingerprint is used for finite -size systems. The entanglement counting develops finite-size effects, which naively have no structure. However, it has been conjectured and numerically substantiated\cite{hermanns11}
that there is a counting principle underlying the finite-size entanglement counting of model states.
Before we turn to the main topic of this paper, namely the PES for the Laughlin states, we mention that the entanglement entropy and spectra are active studied, see references\cite{iblisdir07,rodriguez09,haque09,chandran11,qi12,jackson13,rodriguez13} for some studies in the context of the quantum Hall effect.

We now consider the ground state $| \Psi \rangle$ of a model quantum Hall state, such as the Laughlin\cite{laughlin83}
or Moore-Read\cite{moore91} state, that are the exact zero energy states of a model Hamiltonian.
For a given number  $N$ of particles, this is the unique zero-energy state of a model Hamiltonian
that occurs at the following number of flux quanta
\begin{align}
N_\phi = \frac{1}{\nu} N - \mathcal{S},
\end{align}
where $\nu$ is the filling fraction and $\mathcal{S}$ is the shift. 
Now suppose that the $N$ particles  are divided into two groups, group $A$ containing $N_A$ of the particles, and group $B$ containing the rest 
$N_B=N-N_A$ of them. Without any loss of generality, we can assume that $N_A\le N_B$.
Let \mbox{${\bs x} = ( x_1,x_2, \cdots, x_{N_A} )$} and \mbox{${\bs y} = ( y_1,y_2, \cdots, y_{N_B} )$}
be the coordinates of particles in $A$ and $B$, respectively.
The Schmidt decomposition of the wave function $\Psi ( {\bs z} )$, that is 
\begin{equation}
\Psi ( {\bs z} ) = \sum_{i} e^{-\xi_i/2} \psi^A_i ({\bs x}) \psi^B_i ({\bs y}),
\end{equation}
involves wave functions $\psi^A_i ({\bs x})$ for $N_A$ particles. 
After tracing out the particles of part $B$, we are left with a reduced system of
$N_A$ particles, but the amount of flux remains the same, namely
\begin{align}
N_\phi = \frac{1}{\nu} N_{A} - \mathcal{S} + \Delta N_{\phi},
\end{align}
where $\Delta N_{\phi} = \nu^{-1}N_B$. The presence of this excess flux $\Delta N_{\phi}$ indicates that we should view the reduced system as one with $N_A$ particles, in the presence of quasi-hole excitations. For a {\em real} system with $N_{A}$ particles, and $\Delta N_{\phi}$ excess flux quanta the number of zero-energy states of the model Hamiltonian (which we will call the number of quasi-hole states) can often be obtained exactly\cite{rr96,gr00,arr01,a02,read06}. For instance, in the case of the  $\nu = 1/m$ Laughlin case, quasi-hole states of $N_A$ particles in $\Delta N_{\phi}$ excess flux are of the form
\begin{align}
\psi_i({\bs x}) =  P_i({\bs x}) \prod_{i<j} (x_i-x_j)^m,
\end{align}
where $P_i({\bs x})$ is a symmetric polynomial in $N_A$ variables with degree in each variable at most $\Delta N_{\phi}$. The number of quasi-hole states is therefore
\begin{align}
\binom{\Delta N_{\phi} + N_A}{ N_A}.
\end{align}
This number forms an upper bound for the rank of the reduced density matrix\cite{zozulya07,haque07,chandran11}.

From numerical investigations, it is known that in all cases considered, this upper bound is in fact reached \footnote{Provided no additional trivial constraint, coming from the Hilbert space of the reduced system, is present. The particles of the simplest systems for which this can happen have spin $S=1$.}. This observation has led to the ``rank saturation'' conjecture, which can be thought of as a finite-size version of the Li-Haldane conjecture, namely, \emph{the entanglement level counting of the PES of a model state is equal to the number of bulk quasi-hole states}. This means that the states $\psi^A_i ({\bs x})$ appearing in the Schmidt decomposition of $\Psi ( {\bs z} )$ span all the quasi-hole states of $N_A$ particles in $\Delta N_{\phi}$ excess flux. Proving analytically that this upper bound is indeed reached has proven to be a difficult problem.

In this paper, we revisit this problem for the general $\nu = \frac{1}{m}$ Laughlin states. 
We start by considering the $\nu=1$ Laughlin state, which is simply the Slater determinant
of the completely filled lowest Landau level. We explain how to obtain the rank of the
reduced density matrix of the particle entanglement spectrum in this case. To do so,
we will make some use of the properties of symmetric polynomials. To get a grip on the
$\nu=\frac{1}{m}$ Laughlin states, we then use the following strategy. After
partitioning the particles into two sets $A$ and $B$, we ``split'' the $N_{B}$ particles in part
$B$ into $m N_{B}$ particles, and consider the $\nu =1$ Laughlin state of the system
thus obtained. For this system, we already obtained the rank of the reduced density
matrix. If one can show that clustering the $m N_{B}$ particles into groups of size $m$,
does not lead to a smaller rank of the reduced density matrix,  one deduces the
rank of the reduced density matrix for the $\nu=\frac{1}{m}$ Laughlin state, and shows that
the upper bound is indeed reached.

The hard step in the strategy outlined above is to show that the clustering of the $m N_{B}$
particles into $N_{B}$ groups of of size $m$ does not reduce the rank of the reduced density
matrix. Proving this statement turns out to be highly non-trivial. As we explain in the main text,
one has to show that there is no (non-zero) symmetric polynomial in $m N_{B}$ variables
that vanishes under the formation of $N_{B}$ groups of variables each of size $m$, and whose degree in any of
the variables is $N_{B}$ or less. Although we did not fully succeed in proving this statement,
we did make substantial progress. In particular, we constructed an explicit generating set of the ideal of polynomials that vanish under this clustering. Using this construction we were able to show that  a 
non-zero symmetric polynomial in $m N_{B}$ variables that vanishes under the formation of $N_{B}$
groups of variables each of size $m$ must have a {\em total} degree at least $N_{B}+1$. Proving this weaker
statement is already a non-trivial result, mainly because the positions of the various clusters can
be arbitrary, which means that the clustering condition is non-local. Moreover we were able to prove that all polynomials in the generating set have a degree at least $N_{B}+1$.

The outline of the article is as follows. In section \ref{sec:notation}, we introduce the notion of
partitions, and several types of symmetric polynomials, that we make use of throughout the
article. The PES of the $\nu=1$ Laughlin state is discussed in section \ref{sec:nuisone}.
We continue in section \ref{sec:nuisonem} by explaining how the result for $\nu=1$ can
be used to make progress for the $\nu=1/m$ Laughlin states, and recast the problem in terms
of clustering properties of symmetric polynomials. In section \ref{sec:clustering}, we prove
that the \emph{total} degree of polynomials that vanish under clustering is bounded from below, and
provide an explicit construction for such polynomials in general. In section \ref{sec:possibleproof}, we
make some comments on why it is much harder to prove that not only the total degree,
but also the degree of any variable for polynomials that vanish under the clustering is bounded
from below. In addition, we provide a proof for the statement in the case where one forms two
clusters of size $m$. Finally, we discuss our results in section \ref{sec:discussion}.
In the Appendix A, we derive some properties of the polynomials which are used in section
\ref{sec:clustering}, and in Appendix B we provide an alternate set of polynomials that can be used in
the proof of section \ref{sec:clustering}.

\section{Some notation}
\label{sec:notation}

In this section we introduce some definitions and notations that are used in what follows. We start by introducing the notion of partitions, which play a central role in the theory of symmetric polynomials. For a general introduction to the subject
of partitions, we refer to\cite{book:andrews} and for the theory of symmetric polynomials to\cite{book:macdonald}.

\subsection{Partitions}
\label{sec:partitions}
For a positive integer $\mathcal{D}$, a  non-increasing sequence
$\bs{\lambda}=(\lambda_1,\lambda_2,\dots,\lambda_r)$
of strictly positive integers $\lambda_1$, $\lambda_2$, \dots, $\lambda_r$ is called an
$r$-\emph{partition} of $\mD$ if $\sum_{i=1}^{r}\lambda_i= \mD$.
The $\lambda_i$'s are the \emph{parts} of $\bs{\lambda}$,
and $r$ is called the \emph{length} of $\bs{\lambda}$, which is denoted by $l(\bs{\lambda})$.
We call $\mD$ the \emph{weight} of $\bs{\lambda}$, which is denoted by $|\bs{\lambda}|$.
We write $\bs{\lambda}\vdash \mD$ to indicate that  $\bs{\lambda}$ is a partition of $\mD$.
By convention, $\bs{\lambda}= \emptyset$ is the only partition of zero which we call the empty partition. 
The number of parts of partition $\bs{\lambda}$ which are equal to a given integer $j$ is denoted by $n_j(\bs{\lambda})$ or simply $n_j$.
 We also define 
\eqn{z_{\bs{\lambda}} = \prod_{j=1}^{\lambda_1}j^{n_j} n_j!.}{}

Finally, The set of all partitions of $\mD$ is denoted by $\ptn(\mD)$.
It is not too hard to convince oneself (see Ref.~\onlinecite{book:andrews})
that the number of partitions with at most $r$ parts and  each part at most $d$
is equal to $\tbinom{r+d}{r}$.

\subsection{Symmetric polynomials}
In what comes, we will be dealing with the ring $\Lambda_N$ of \emph{symmetric polynomials} in $N$ variables.
A polynomial $P$ is called a symmetric polynomial in $N$ variables, if for all permutations $\sigma$ of $\{1,\ldots,N\}$,
\eqn{
P\l(x_{\sigma(1)},\dots,x_{\sigma(N)}\r)=P(x_1,\dots,x_N).
}{e000}
The degree $d$ of a symmetric polynomial is simply the degree in one of its variables.

A polynomial $P(x_1,\dots,x_N)$ is called homogeneous of total degree $\mathcal{D}$, if for any real number $l$,
\begin{align}
P( l \, x_1,\dots, l \, x_N) = l^{\mathcal{D}} P(x_1,\dots,x_N).
\end{align}
For instance, the polynomial $P(x_1,x_2) = x_1^2x_2 + x_1x_2^2 $ is a homogeneous symmetric polynomial of degree \mbox{$d=2$} and total degree $\mathcal{D}=3$. 
 
There are different bases that one can consider for $\Lambda_N$. A natural one,  is given by the set of, so-called, \emph{symmetric monomials}.
Given a partition $\bs{\lambda}=(\lambda_1,\lambda_2,\cdots,\lambda_r)$ with $r \leq N$,
the symmetric monomial  $m_{\bs{\lambda}}(x_1,\cdots,x_N)$ is defined as
\begin{widetext}
\begin{align}
m_{\bs{\lambda}}(x_1,\cdots,x_N):= \sum_{\sigma}{} 
x_{\sigma(1)}^{\lambda_1}  x_{\sigma(2)}^{\lambda_2} \cdots  x_{\sigma(r)}^{\lambda_r}  x_{\sigma(r+1)}^{0} \cdots x_{\sigma(N)}^0,
\label{e001}
\end{align}
\end{widetext}
where the sum is over all \emph{distinct} permutations $\sigma$ of the parts of $\bs{\lambda}$,
and it is defined to be $1$ if $\bs{\lambda}$ is the empty partition.
On the other hand if $r >N$ we set $m_{{\bs \lambda}} (x_1,\cdots ,x_N) = 0$. For example,
\begin{widetext}
\eqn{
m_{(2,1,1)}(x_1,x_2,x_3)&=
x_1^2x_2x_3+x_1x_2^2x_3+x_1x_2x_3^2,\\
m_{(2,1)}(x_1,x_2,x_3)&
=x_1^2x_2+x_1x_2^2+x_1^2x_3+x_1x_3^2+x_2^2x_3+x_2x_3^2,\\
m_{(2,1,1)}(x_1,x_2)&=0.}{e002}
\end{widetext}
When studying rank saturation of the PES for the Laughlin state, finite-size effects imply an upper bound for the degree of polynomials. We will therefore be led to consider the space $\Lambda_N^d$ of symmetric polynomials in $N$ variables, with degree (in each of the variables) at most $d$. A basis for this space is given by the symmetric monomials $m_{{\bs \lambda}}(x_1,\ldots,x_N)$ corresponding to partitions $\bs{\lambda}$ with at most $N$ parts and each part at most $d$. Therefore, we have
\begin{align}
\dim \left(\Lambda_N^d \right) = \binom{N+d}{N}.
\end{align}

Another important family of symmetric polynomials is the set of \emph{elementary symmetric polynomials}.
The elementary symmetric polynomials that are labelled by an integer $n$ are defined in terms of symmetric monomials as
$e_n:=m_{(\underbrace{1,\dots,1}_{\text{$n$ ones}})}$. For instance,
\begin{align}
e_{0} (x_1,x_2,x_3) &= 1, \nonumber\\
e_{1} (x_1,x_2,x_3) &= x_1 + x_2 + x_3, \nonumber\\
e_{2} (x_1,x_2,x_3) &= x_1 x_2 + x_1 x_3 + x_2 x_3, \nonumber\\
e_{3} (x_1,x_2,x_3) &= x_1 x_2 x_3, \nonumber\\
e_{n\geq 4} (x_1,x_2,x_3) &= 0.
\end{align}
For a partition \mbox{$\bs{\lambda}=(\lambda_1,\ldots,\lambda_r)$}, the elementary symmetric polynomial
$e_{\bs{\lambda}}$ is defined as \mbox{$e_{\bs{\lambda}}:=e_{\lambda_1}\cdots e_{\lambda_r}$}.
As an example,
\eqn{
e_{(2,1,1)}(x_1,x_2)&=e_2(x_1,x_2)e_1(x_1,x_2)e_1(x_1,x_2)\\&=
x_1x_2(x_1+x_2)^2.
}{e003}
It is known that the set of all polynomials $e_{\bs{\lambda}}(x_1,\ldots,x_N)$, where $\bs{\lambda}$ is a partition with at most
$d$ parts and each part at most $N$, forms a basis of the space $\Lambda_N^d$.

Lastly, the \emph{power sum symmetric polynomials}, defined as  
\begin{align}
p_i(x_1,\ldots,x_N) :=  x_1^i + \cdots + x_N^i  \, ,
\end{align}
are of special importance. In fact, the set $\{p_1, p_2, \ldots, p_N \}$ generates $\Lambda_N$. This means that any symmetric polynomial $P$ in $N$ variables can be written as a
polynomial in $(p_1,\ldots,p_N)$. In other words, the set of polynomials $p_{\bs{\lambda}} := p_{\lambda_1} \cdots p_{\lambda_r}$,
where $\bs{\lambda}$ is a partition with each part at most $N$, forms a basis of $\Lambda_N$.
For example,
\begin{align}
e_2 = \sum_{i<j} x_i x_j = \frac{p_1^2-p_2}{2},
\end{align}
independent of the number of variables $N$.
Most importantly, the decomposition of any symmetric polynomial $P$ in $N$ variables as a polynomial
in $(p_1,\ldots,p_N)$ is unique. One should note that, unlike for the symmetric monomials $m_{\bs{\lambda}}$ and the elementary symmetric polynomials $e_{\bs{\lambda}}$, there is no natural restriction on ${\bs \lambda}$ such that the corresponding $p_{\bs{\lambda}}$'s form a basis for $\Lambda_N^d$.

\section{The $\nu=1$ state}
\label{sec:nuisone}

To obtain the rank of the reduced density matrix of the Laughlin states in the case of
the ``particle cut'', we start by considering the simplest case, the $\nu=1$ Laughlin state,
which is just a single Slater determinant,
\begin{equation}
\Psi_{\nu=1}(z_1,\ldots,z_N)=
\prod_{1 \leq i<j \leq N}(z_i-z_j),
\label{eq:nu1}
\end{equation}
up to a geometry-dependent Gaussian factor. For instance, the plane and sphere geometry give rise to different Gaussian factors, inherited from the respective inner products. However, for our purposes the precise form of the Gaussian factor is irrelevant. The results presented in this paper involves only the notion of linear independence, and does not refer to the notion of orthogonality. As a consequence, the underlying inner product plays no role and our result is equally valid on the plane, sphere, and cylinder.

Now suppose that the $N$ particles  are divided into two groups $A$, containing $N_A$ of the particles, and $B$ containing $N_B=N-N_A$ particles. At this stage we do not assume $N_A\le N_B$. Let us rename the coordinates of particles in $A$ and $B$ to ${\bs x} = (x_1,x_2, \cdots, x_{N_A})$ and ${\bs y} = (y_1,y_2, \cdots, y_{N_B} )$, respectively.  The rank of the reduced density matrix in the case of such a particle cut can be obtained from a decomposition of the wave function $\Psi ( {\bs z} ) $ of the form
\begin{equation}
\Psi ( {\bs z} ) = \sum_{i}  \psi^A_i ({\bs x}) \psi^B_i ({\bs y}), \label{pseudo Schmidt}
\end{equation}
where the set of wave functions  $\psi_i^A$ (resp. $\psi^B_i$) are independent. Note that this is not quite a Schmidt decomposition since we do not demand the $\psi_i^A$'s to form an orthonormal set. Although this is not a Schmidt decomposition, the number of terms in the sum is equal to the Schmidt rank, or equivalently, to the rank of the reduced density matrix. Therefore, we will call the decomposition \eqref{pseudo Schmidt} a Schmidt decomposition, although strictly speaking this is an abusive notation.

Before we explicitly write the $\nu=1$ Laughlin state in such a ``Schmidt-decomposed" form,
we note that we can obtain the rank of the reduced density matrix in the $\nu = 1$ case in
a straightforward way. This state is simply obtained by filling the Landau orbitals from $0$ up to $N_{\Phi} = N-1$
\begin{align}
| \Psi_{\nu=1} \rangle = | 111 \cdots 111 \rangle.
\end{align}
The Schmidt decomposition relative to particle cut amounts to choose $N_A$ out of the $N$ particles
\begin{widetext}
\begin{multline}
| \Psi_{\nu=1} \rangle   \propto  | \underbrace{111 \cdots 11}_{N_A} 00 \cdots 0 \rangle \otimes | 000 \cdots 0011 \cdots 1 \rangle\\ +  | \underbrace{111 \cdots 10}_{N_A} 10 \cdots 0 \rangle \otimes | 000 \cdots 0101 \cdots 1 \rangle  + \cdots \\+ | \underbrace{000 \cdots 00}_{N_B} 11 \cdots 1 \rangle \otimes | 111 \cdots 1100 \cdots 0 \rangle,
\end{multline}
\end{widetext}
which means that the rank of the reduced density matrix is given by the number of ways in which the
$N_A$ particles of system $A$ can be divided over the number of orbitals.
The number of orbitals is given by $N = N_{A} + N_{B}$, so we obtain that the rank of
the reduced density matrix is given by $\tbinom{N_A + N_B}{N_A}$. 
We note that the same result can be obtained  directly from the wave function\cite{sterdyniak12},
which is a single Slater determinant
$\Psi_{\nu=1} ({\bs z} )  = \prod_{i <j} (z_i-z_j)$.

\begin{figure}[ht]
\includegraphics[width=7cm]{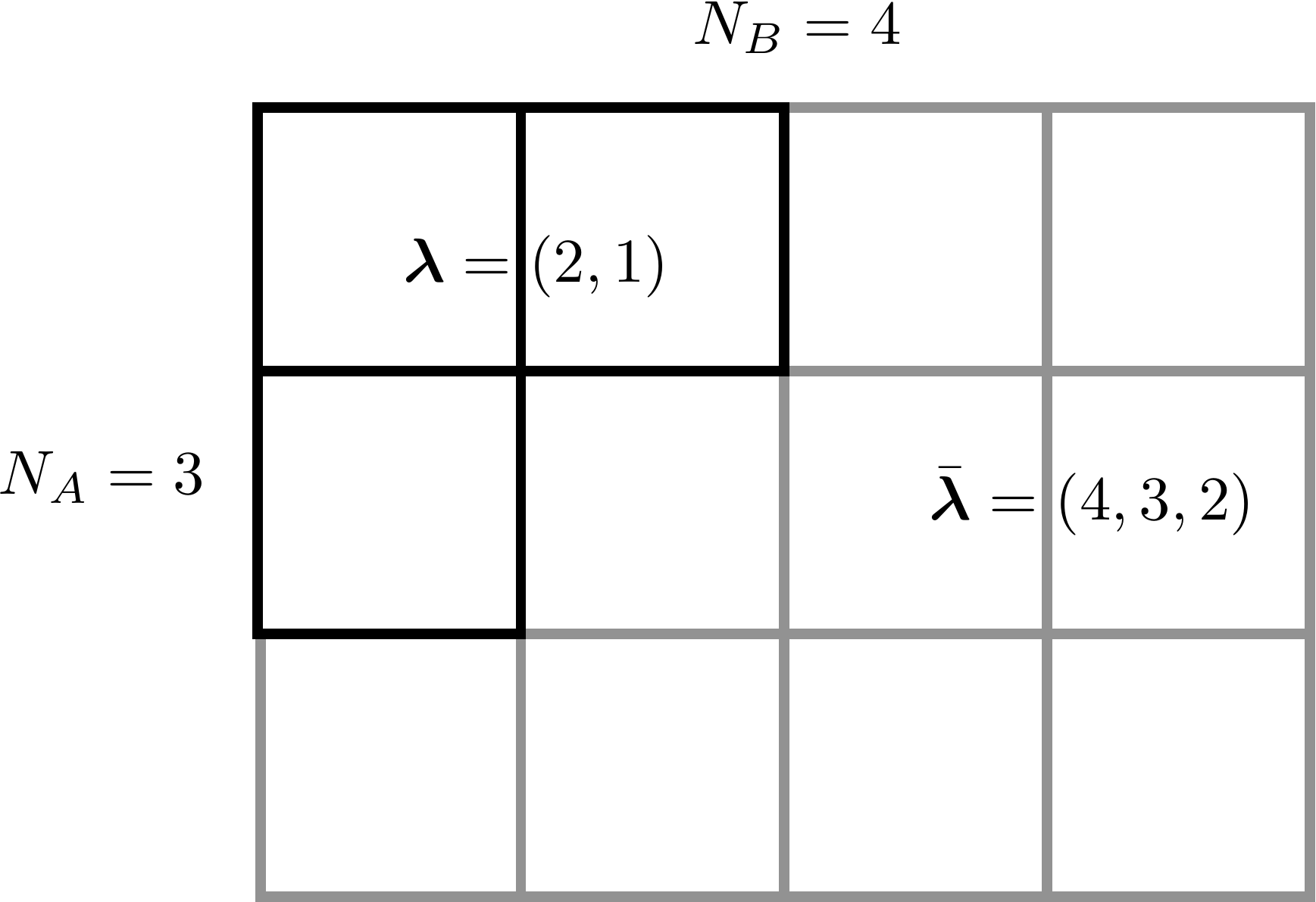}
\caption{The relation between the partition $\bs\lambda$ and its complement $\bar{\bs \lambda}$ for given $N_A$ and $N_B$.}
\label{fig:lambda-bar}
\end{figure}

It is instructive to perform a more explicit  Schmidt decomposition of the $\nu=1$ Laughlin wave function.
We start by writing the state explicitly in terms of the
variables $x_i$ and $y_i$ of groups $A$ and $B$, respectively.
Dropping the exponential factors, we have
\begin{equation}
\Psi_{\nu=1} ({\bs z} )  = \Psi_{\nu=1} ({\bs x} )\l( \prod_{i,j} (x_i - y_j)\r)  \Psi_{\nu=1} ({\bs y} ). 
\end{equation}
We are going to use the following result \cite{book:macdonald}
\begin{equation}
\prod_{i=1}^{N_A} \prod_{j=1}^{N_B} (1+ x_i y_j) =
\sum_{{\bs \lambda}} m_{{\bs \lambda}} ({\bs x}) e_{{\bs \lambda}} ({\bs y}).
\label{eq:me-formula}
\end{equation}
where the sum is over all partitions ${\bs \lambda}$ with maximally $N_A$ parts, and each part being maximally $N_B$, i.e., all partitions which fit in a rectangle of height $N_A$ and width $N_B$. Thus,
\begin{align}
 \prod_{i,j} (x_i - y_j)  = \sum_{{\bs \lambda}} (-1)^{|{\bs \lambda}|} m_{\bar{\bs \lambda}}({\bs x}) e_{{\bs \lambda}} ({\bs y}).\label{Theta_decomposition}
\end{align}
Here, we used the relation
\begin{align}
\left( \prod_i x_i^{N_B} \right)  m_{\bs \lambda} (-1/{\bs x}) = (-1)^{|{\bs \lambda}|} m_{\bar{\bs \lambda}}({\bs x}),
\end{align}
where the partition $\bar{\bs \lambda}$ stands for the complement of ${\bs \lambda}$ with respect to the rectangle of height $N_A$ and width $N_B$. In addition, $(-1/{\bs x})$ is shorthand for
$(-1/x_1,\ldots,1/x_{N_A})$.
As an example, it is shown in Fig.~\ref{fig:lambda-bar}  that
for $N_A = 3$, $N_B = 4$ and $\bs\lambda = (2,1)$, one finds $\bar{\bs \lambda} = (4,3,2)$.
We then obtain a  Schmidt decomposition for the $\nu= 1$ Laughlin state
\begin{equation}
\Psi_{\nu=1}(\bs z) = \sum_{{\bs \lambda}}   (-1)^{|{\bs \lambda}|}  q^A_{{\bs \lambda}} ({\bs x}) q^B_{{\bs \lambda}} ({\bs y}),
\end{equation}
where 
\eqn{
q^A_{{\bs \lambda}} ({\bs x})  =   m_{\bar{\bs \lambda}}({\bs x})\,   \Psi_{\nu=1} ({\bs x}),
}{}
and
\eqn{
q^B_{{\bs \lambda}} ({\bs y})=e_{{\bs \lambda}} ({\bs y})\, \Psi_{\nu=1} ({\bs y}).
}{} 
A few remarks on this formula are in order here. The number of terms in the sum
is most important here. The sum over ${\bs \lambda}$, is over all
partitions with maximally $N_A$ parts, and each part being maximally $N_B$,
i.e., all partitions which fit in a rectangle of height $N_A$ and width $N_B$.
There are $\binom{N_A + N_B}{N_A}$ such partitions. The rank of
the reduced density matrix of the $\nu =1$ Laughlin state is thus given by
\begin{align}
\binom{N_A + N_B}{N_A}, 
\end{align}
and we recover the dimension of $\Lambda_{N_A}^{N_B}$. It is straightforward to check that this is also the dimension of the set of anti-symmetric polynomials in $N_A$ variables with maximum degree $N = N_A + N_B$, which is nothing but the space of `quasi-hole' states for the non-interacting $\nu=1$ case. The set of polynomials $q^A_{{\bs \lambda}}$  (resp. $q^B_{{\bs \lambda}}$) forms a basis for  the space of anti-symmetric polynomials in $N_A$ (resp. $N_B$) variables with maximum degree $N$.  Note that this result is symmetric under exchange of $A$ and $B$, and in particular it holds whether or not $N_A \leq N_B$. This is a particularity of the $\nu =1$ case and it will no longer be true for the $\nu=\frac{1}{m}$
Laughlin state with $m>1$.

\section{Schmidt decomposition of the $\nu= \frac{1}{m}$ Laughlin state}
\label{sec:nuisonem}

We are now going to compute the rank of the reduced density matrix for the generic $\nu = 1/m$ Laughlin state
\begin{align}
\Psi_{m}(z_1,\ldots,z_N)=\prod_{1\le i<j\le N}(z_i-z_j)^m.
\end{align}
As usual we divide the particles into two groups $A$ and $B$, containing $N_A$ and $N_B=N-N_A$ of them, respectively, and  we assume that \mbox{$N_A\le N_B$}.  We are interested in obtaining a  Schmidt decomposition of this state.
As for the $\nu=1$ case, we can write
\begin{equation}\label{EQ33}
\Psi_{m} ({\bs z} )  = \Psi_{m} ({\bs x} )\l(\prod_{i=1}^{N_A}\prod_{j=1}^{N_B}(x_i-y_j)^m \r) \Psi_{m} ({\bs y}).
\end{equation}
Proving that the rank of the PES for the $\nu=1/m$ Laughlin state is saturated boils down to finding a Schmidt decomposition for the wave function (\ref{EQ33}) and counting the number of terms in the decomposition. As in the $\nu=1$ case, one need to take care of only the middle term of the wave function (\ref{EQ33}) 
\eqn{
\Phi_{m}(\bs{x},\bs{y}):=\prod_{i=1}^{N_A}\prod_{j=1}^{N_B}(x_i-y_j)^m.
}{eq5}
To do so we start from
\eqn{\Phi_1 (\bs{x},\bs{w})=\prod_{i=1}^{N_A}\prod_{j=1}^{mN_B}(x_i-w_j),}{}
for $\bs{w}=(w_1,\ldots,w_{mN_B})$. From \eqref{Theta_decomposition},
\eqn{
\Phi_1 (\bs{x},\bs{w})=\sum_{{\bs \lambda}} (-1)^{|{\bs \lambda}|} m_{\bar{\bs \lambda}}({\bs x}) e_{{\bs \lambda}} ({\bs w}),
}{Theta}
but this time the sum is over all partitions $\bs{\lambda}$ which fit in a rectangle of height $N_A$ and width $mN_B$. 
We then relate $\Phi_{m}$ to $\Phi_1$ through the clustering transformation, which is a linear transformation from $\Lambda_{m N_B}$ to $\Lambda_{N_B}$ defined as follow. 

To a symmetric polynomial $P({\bs w})$ in $m N_B$ variables,  we associate the polynomial 
variables $\mC_m \bigl(P ({\bs y})\bigr)$ in $N_B$ variables obtained by regrouping the particles into clusters of $m$, i.e.,
\begin{widetext}
\begin{align}
(\mC_m P) (y_1,y_2,\ldots,y_{N_B}) = P( \underbrace{y_1,y_1,\ldots,y_1}_{m},\underbrace{y_2,y_2,\ldots,y_2}_{m},\ldots,\underbrace{y_{N_B},y_{N_B},\ldots,y_{N_B}}_{m} ). \label{clustering}
\end{align}
\end{widetext}
It is easy to see that after clustering $\Phi_1$ becomes $\Phi_{m}$, i.e.,
\eqn{
\mC_m (\Phi_1) =\Phi_{m}.
}{eq9}
Applying the clustering transformation to both sides of Eq.~(\ref{Theta}) results in
\begin{align}
\Phi_{m} (\bs{x},\bs{y})= \sum_{{\bs \lambda}}  (-1)^{|{\bs \lambda}|}m_{{\bs \lambda}} ({\bs x}) \mC_m \bigl(e_{{\bs \lambda}}\bigr) ({\bs y}).
\end{align}
As mentioned earlier, the sum  is over all 
partitions ${\bs \lambda}$ with maximally $N_A$ parts, and each part being maximally $ m N_B$.
There are $\binom{N_A + m N_B}{N_A}$ such partitions. This is precisely the number of Laughlin quasi-hole states for $N_A$ particles in $\Delta N_{\Phi} = m N_B$ extra fluxes, and we recover the usual upper bound for the rank of the reduced density matrix. 

Rank saturation of the PES for the $\nu = 1/m$ Laughlin state boils down to the following, non-trivial result:
the polynomials $\mC_m ( e_{\bs \lambda})$ are independent.
More precisely, one has to prove that the linear transformation 
\eq{
\Lambda_{m N_B}^{N_A}&\longrightarrow\Lambda_{N_B}^{m N_A}\\P(\bs{w})&\longrightarrow\mathcal{C}_m(P)(\bs{y})
}{}
is injective as long as $N_A \leq N_B$. Since $\dim \Lambda^{ m N_A}_{N_B} \geq \dim \Lambda_{ m N_B}^{N_A} $,
it is sufficient to show that this linear map has a trivial kernel. Namely, besides $P=0$, no polynomial in $m N_B$ variables
and maximum degree $N_A$ can vanish under the clustering transformation.

\section{Clustering Properties of Symmetric Polynomials}
\label{sec:clustering}

In this section we are going to describe the ideal of symmetric polynomials in $q = m\, N$ variables that vanishes under the clustering transformation \eqref{clustering}. In particular,  we are going to construct a generating set of this subspace, and prove that a non-zero symmetric polynomial in $q= m\,N$ variables that vanishes under the clustering transformation has a {\em total} degree $\mD$ of at least $N+1$. We are also going to prove that this symmetric polynomial of minimal total degree is unique (up to a scaling numerical factor).

The statement that the total degree of a symmetric polynomial in $q= m\,N$ variables that vanish
under the clustering conditions is at least $N+1$ is a weaker
statement than stating that the degree of each variable is at least $N+1$, but easier to
prove. After finishing the proof of the statement on the total degree, we comment on how
one might prove the stronger statement, limiting the degree of the polynomials.

As a warmup, we start with two simple examples, which
we will come back to after the proof. We start with the case $m=2$ and $N=1$, i.e., we are looking
for a symmetric polynomial in two variables $y_1$ and $y_2$,  of total degree $2$,  that vanishes when $y_1=y_2$. It is
easy to see that the polynomial has degree two, namely $(y_1-y_2)^2$.

The case $m=2$ and $N=2$ is already more complicated. With some thought, one can construct
a total degree $3$ symmetric polynomial in four variables $y_1,\ldots,y_4$, that vanishes when $y_1=y_2$ and
$y_3 = y_4$, namely $(y_1+y_2-y_3-y_4)(y_1-y_2+y_3-y_4)(y_1-y_2-y_3+y_4)$. It is already less trivial
to convince oneself that no lower degree symmetric polynomial with the same vanishing properties
exists. Upon increasing the $m$ and $N$, even finding polynomials with the correct vanishing
properties becomes a hard problem, which is caused by the non-locality of their defining property. Namely,
polynomials have to vanish, independent of the position of the various clusters.
As we indicated above, we solve this problem in a constructive way.

Our construction is motivated by the following observation. The ring $\Lambda_q$ of symmetric polynomials in $q = m\, N$ variables is generated by $\{ p_1, \ldots, p_q\}$, and the power sum polynomials $p_i$ have a very simple behavior under the clustering transformation \eqref{clustering}, namely 
\begin{align}
\mC_m (p_{i})  = m \, p_{i}.
\end{align}
However, after the clustering transformation there are only $N$ variables left. This means that \mbox{$\mathcal{A}_N =\{p_{1},\ldots,p_{N}\}$} forms  a minimal set of generators, and the polynomials $\{ p_{N+1}, \cdots, p_q \} $ are no longer independent after being clustered. The generators $p_i$ are not very convenient to describe the clustering transformation, and this is why we introduce a new set of generators $\tilde{\mathcal{A}}_q=\{\tilde{p}_{1},\ldots,\tilde{p}_{q}\}$ of $\Lambda_q$ as 
\begin{align}
\tilde{p}_n =  \sum_{ \bs{\mu}\vdash n}  (-1)^{|\bs{\mu}|} \left(- \frac{1}{m}\right)^{l(\bs{\mu})} \frac{ p_{\bs{\mu}}}{z_{\bs\mu}}\cdot 
\end{align}
Alternatively, the polynomials $\tilde{p}_n $ can be defined in terms of the polynomials $r_n^{(x)}$ of Appendix \ref{Polynomials r_n} through $\tilde{p}_n =  \frac{(-1)^n}{n!}  r_n^{(1/m)} $. The main property of these new polynomials is that they behave nicely under clustering:
 \begin{align}
\mC_m \bigl( \tilde{p}_{n})  & =e_n, \qquad  n =1 ,\ldots, N, \label{ptilde_property1} \\
\mC_m \bigl( \tilde{p}_{n})  & =0, \qquad  n  > N, \label{ptilde_property2}
\end{align}
as inherited from the properties of $r_n^{(x)}$ described in Appendix \ref{Polynomials r_n}.
We also introduce the $\tilde{p}_\lambda$ in the usual way as
$\tilde{p}_{\bs\lambda} = \tilde{p}_{\lambda_1} \ldots  \tilde{p}_{\lambda_r}$, where $r$ is the length of
$\bs\lambda$.
In terms of these modified power sums $\tilde{p}_n$,  it is now relatively easy to describe the ideal of polynomials in $\Lambda_q$ that vanish under the clustering transformation \eqref{clustering}: 

\thm{
The set $\{ \tilde{p}_{\bs\lambda} | N+1 \leq \lambda_1 \leq q \}$ forms a basis of the ideal of
symmetric polynomials in $q=mN$ variables that vanish under the clustering transformation.
In other words, this ideal is generated by the set
$\{ \tilde{p}_{N+1},\tilde{p}_{N+2}, \ldots, \tilde{p}_{q} \}$.}{}
\prf{
Suppose that $P$ is a symmetric polynomial in $q$ variables. Because $\tilde{\mathcal{A}}_q$ is
a generating set, there exists a polynomial $R$ in $q$ variables such that
\eq{
P = R(\tilde{p}_{1},\ldots,\tilde{p}_{q}).
}{}
Generically, there are two kinds of monomials in the polynomial $R$. Those that depend only on the first $N$ variables $\tilde{p}_{1},\ldots,\tilde{p}_{N}$, and the ones that depend on at least one of the $\tilde{p}_{n}$, with $n>N$. Accordingly, $R$ can be decomposed uniquely into a sum of two polynomials
\eq{
R(\tilde{p}_{1},\ldots,\tilde{p}_{q}) = A(\tilde{p}_{1},\ldots,\tilde{p}_{N})+ B(\tilde{p}_{1},\ldots,\tilde{p}_{q}).
}{}
Thus, by construction, $\mC_m \bigl( B \bigr) = 0$.
It is now straightforward to check that $\mC_m \bigl( R \bigr) = 0$ if and only if $A=0$, since 
\eq{
\mC_m \bigl( P \bigr) = A( e_{1} ,\ldots, e_{N}), 
}{}
and the $\{e_{1} ,\ldots, e_{N}\}$ are algebraically independent in $N$ variables.
Therefore the set 
$\{ \tilde{p}_{\bs\lambda} | N+1 \leq \lambda_1 \leq q \}$
is a basis of the kernel of the clustering transformation.}

\cor{The only symmetric polynomial $P$ in $mN$ variables with total degree $N$ or less that vanishes under the clustering transformation is $P=0$. Moreover, $\tilde{p}_{N+1}$ is the unique (up to an overall factor) symmetric polynomial in $q$ variables and total degree $N+1$ that vanishes under  this clustering.}{cor001}

Since the modified power sum $\tilde{p}_n$ has total degree $n$, this corollary follows directly from Theorem 1.  Let us illustrate this with two simple examples.

\ex{Consider the simplest non-trivial case where $q=2$, $N=1$, and $m=2$. In this case,
the clustering condition is $y_1=y_2$. The definition of $\tilde{p}_2$ yields
\begin{align}
\tilde{p}_2 = \frac{1}{8}\left( p_1^2 - 2 p_2 \right) = -\frac{1}{8}(y_1-y_2)^2,
\end{align}
which reproduces the expected result.
}{}

\ex{As another example, let $q=4$, $N=2$, and $m=2$. This time, clustering conditions 
are $y_1=y_2$ and   $y_3=y_4$. We have 
\begin{align}
\tilde{p}_3 = \frac{1}{48}( p_1^3 - 6 p_2 p_1 + 8 p_3).
\end{align}
For $q=4$ variables, this is 
\begin{align}
\tilde{p}_3 = \frac{1}{16}(y_1+y_2-y_3-y_4)(y_1-y_2+y_3-y_4) (y_1-y_2-y_3+y_4).
\end{align}
}{}

We should note that these two examples are not representative for the general case,
in the sense that the polynomials $\tilde{p}_{N+1}$ do not generically factorize to a
simpler form. For instance, for $N=3$ and $m=2$, we have
\eq{\tilde{p}_{4} =
\frac{1}{384}(p_{1}^4-12 p_2p_{1}^2+12p_{2}^2+32p_{1} p_{3}-48 p_{4}),}{}
which does not have a simple factorized form when restricting to $q=6$ variables.

\con{There is no non-zero symmetric polynomial $P$ in $mN$ variables with degree $N$ or less that vanishes under the clustering transformation.}{conj strong}

While we know that the modified power sum $\tilde{p}_n$ has degree $n$ (see Appendix \ref{Polynomials r_n}), this is not sufficient to prove this conjecture. 

\section{$SU(2)$ invariance}
\label{sec:su(2)}

In the context of the fractional quantum Hall effect, there is a natural action of $SU(2)$ on $\Lambda_{N}^{N_{\Phi}}$ coming from the rotational invariance of the sphere.  The angular momentum operators on the sphere\cite{h83} are
\begin{align}
L^-  & = \sum_{i=1}^N \frac{\partial}{\partial z_i}, \\
L^3  & = \sum_{i=1}^N  \left( z_i \frac{\partial}{\partial z_i}  - \frac{N_{\Phi}}{2} \right), \\
L^+  & = \sum_{i=1}^N \left( z_i N_{\Phi} - z_i^2 \frac{\partial}{\partial z_i} \right). 
\end{align} 
Every polynomial $P$ in $\Lambda_{N}^{N_{\Phi}}$ has a $SU(2)$ symmetric $\Omega(P)$ with opposite angular momentum $L^3$ given by 
\begin{align}
\Omega(P)(z_1,\cdots,z_N) = \left(\prod_{i=1}^N z_i^{N_{\Phi}}\right) P(1/z_1,\cdots, 1/z_N).
\end{align}
Under this $\mathbb{Z}_2$ operation, $L^-$ and $L^+$ are exchanged   and $L^3 \to -L^3$. 

These linear operators are compatible with the clustering, in the sense that $\mC_m L^i = L^i \mC_m$. Note that in these identities the $SU(2)$ operators in the l.h.s. act in $\Lambda_{m N}^{N_{\Phi}}$, while in the r.h.s. they act in $\Lambda_{N}^{m N_{\Phi}}$:
\begin{align}
\begin{CD}
\Lambda_{m N}^{N_{\Phi}}  @> L^i>>  \Lambda_{m N}^{N_{\Phi}} \\
@VV\mC_mV        @VV\mC_mV\\
\Lambda_{N}^{ m N_{\Phi}}     @> L^i>>  \Lambda_{N}^{m N_{\Phi}}
\end{CD}
\end{align}
The same is true for the $\mathbb{Z}_2$ operation $\Omega$. These commuting properties are straightforward to check for $L^3$, $L^-$, and for $\Omega$. Therefore, it also holds for $L^+ = \Omega L^- \Omega$.  For instance, the clustering transformation clearly preserves the total degree, hence the action of clustering commutes with $L^{3}$. Likewise, $L^-$ being the generator of  global translations, it commutes with the clustering.  The following theorem follows immediately:

\thm{The ideal of symmetric polynomials in $mN$ variables that vanish under the clustering transformation is invariant under the action of $L^i$ and $\Omega$.}{thm su2}

\cor{The polynomial $\tilde{p}_{N+1}$ is translationally invariant.}{}
The polynomial $\tilde{p}_{N+1}$ vanishes under clustering, and therefore so does $L^-\tilde{p}_{N+1}$. If this last polynomial was non-zero, it would have a total degree $N$, which is forbidden by Corollary 1. Therefore $L^-\tilde{p}_{N+1} = 0$ and $\tilde{p}_{N+1}$ is translationally invariant.

In fact, it is possible to directly calculate $L^{-} \tilde{p}_{i}$, and one gets
\begin{align}
L^{-} \tilde{p}_{i} = (N +1-i)\tilde{p}_{i-1}.
\label{eq:Lminptilde}
\end{align}
Note that this results only hold for $q=mN$ variables. This follows from the behavior of $r_n^{(x)}$ under translations, which is given in Appendix~\ref{app:ptildederivative}.

Since the kernel of $\mC_m$ is invariant under the action of $L^i$, it can be decomposed into irreducible representations of $SU(2)$. In order to prove that non zero polynomials that vanish under the clustering have degree at least $N+1$, it is therefore sufficient to prove it for lowest weights, that is to say translation invariant polynomials. Therefore Conjecture \ref{conj strong} is equivalent to the following:

\con{The only translationally invariant symmetric polynomial $P$ in $mN$ variables with degree $N$ or less that vanishes under the clustering transformation is $P=0$.}{conj weak}

\section{A possible road towards finishing the proof}
\label{sec:possibleproof}

As we saw in the previous section, we were able to prove that the {\em total} degree of
a symmetric polynomial is at least $N+1$, if the polynomial vanishes under the clustering
transformation \eqref{clustering}. However, we would like to show that the the maximum degree of
any of the variables (i.e., the number of fluxes $N_\phi$) is at least $N+1$. Proving this
statement turns out to be much harder than it looks at first. One of the reasons is
that the clustering we consider is a non-local process. Namely, the positions of the various
clusters are arbitrary. Therefore, proving that the total degree is bounded from below is already
a non-trivial result. What makes proving a bound on the total degree more tractable in
comparison to proving a bound on the degree, is that upon taking linear combinations,
the total degree of the polynomials does not change, provided the resulting polynomial
does not vanish. The degree of the polynomial, however, can be lowered by taking linear
combinations. 

To show that the rank of the reduced density matrix for the particle cut does indeed satisfy
the upper bound given in the introduction, it suffices to prove that the clustering map
$\mC_m: \Lambda_{m N_B}^{N_A} \longrightarrow \Lambda_{N_B}^{m N_A}$, is
injective if $N_A \leq N_B$. In the case $N_A = N_B$, the map $\mC_m$ would then actually
be bijective. One possible route in trying to prove this, is to find two suitable bases for
$\Lambda_{mN}^{N}$ and $\Lambda_{N}^{mN}$ in which 
the map $\mC_m$ acts in an upper-triangular way, and then check that all diagonal elements are non-zero. We did not, however, succeed in finding suitable bases.

A completely different route to prove that the rank of the reduced density matrix is given
by the upper bound is to try to make use of the results for the Read-Rezayi states\cite{rr99}.
These states are defined by the property that they vanish if $k+1$ particles are put at the
same location (in their simplest bosonic incarnation). In particular, it is known exactly
 that how many symmetric polynomials there are, that satisfy this clustering condition,
for an arbitrary number of particles, and arbitrary degree\cite{a02,read06}.
In addition, there are explicit expressions for these polynomials\cite{read06},
see also\cite{aks05}. Using these results, we can prove the wanted result for $N=2$ and
arbitrary $m$. That is, we can show that any symmetric polynomial in $2m$ variables,
that vanishes if two clusters of $m$ variables each are formed, has degree at least three.

To do so, assume that $P$ is a polynomial in $2m$ variables that vanishes under
the clustering, $\mC_m P = 0$. We know that the
total degree of this polynomial is at least 3, and
we want to show that the minimal value of the degree is three as well.

To show this, we note that the polynomial $P$ also vanishes if we make
one big cluster of $2m$ variables. From the results on the Read-Rezayi states,
we know that such symmetric polynomials have degree at least two
(it vanishes, so it should vanish quadratically), and
that it is unique (up to an overall factor). In addition, we know an explicit form
of this polynomial $P'$, namely
\begin{equation}
P' (z_1,\ldots,z_{2m}) = S[ (z_1 - z_3)(z_2 - z_3) ],
\end{equation}
where $S$ denotes the complete symmetrization over all $2m$ variables.
In this case,  by inspection, one can convince oneself that $P'$ does not vanish
if one makes two clusters of $m$ variables. Thus, the minimal degree of a
polynomial $P$ in $2m$ variables that does vanish under $\mC_m$ has degree
at least three. In fact, it is not too hard to find an expression similar to the one
for $P'$, namely
\begin{equation}
P (z_1,\ldots,z_{2m}) = S[ (z_1 - z_4)(z_2 - z_4)(z_3 - z_4) ].
\end{equation}
It is not completely obvious that this vanishes under the clustering for $m>2$,
but one can convince oneself that after symmetrization, one indeed does get zero.

Though it is not going to be easy, one could try to proceed in this way. Constructing the next case, namely polynomials that vanish for three clusters of $m$ variables,
is already more involved. Writing down an explicit form similar to the ones above
is not straightforward, but one can for instance symmetrize the following combination
\begin{widetext}
\begin{equation}
\label{eq:polmis3}
P (z_1,\ldots,z_{3m}) =
S[ (z_1 - z_5)(z_2 - z_5)(z_3 - z_5)(z_4 - z_5) - (z_1 - z_5)(z_2 - z_5)(z_2 - z_6)(z_3 - z_6) ].
\end{equation}
\end{widetext}
This polynomial is the unique polynomial (up to a constant factor)
in $3m$ variables, of degree and total degree $4$, that vanishes under formation of three
clusters of $m$ variables. We stress, however, that this alone does not imply that there
are no polynomials of degree three, that vanish under the same clustering conditions.

The lowest degree polynomial for $N=4$ and arbitrary $m$ can still be written by
symmetrizing an expression like the one in Eq.~\eqref{eq:polmis3}, i.e., two terms only,
but it seems likely that these expressions become more complicated upon increasing $N$.
In addition, having these explicit expressions does not help in excluding the existence of
lower degree polynomials with the same clustering conditions.

\section{Discussion}
\label{sec:discussion}

In this paper, we revisited the study of the PES for the $\nu=1/m$ Laughlin states, in particular
the rank of the associated reduced density matrix. To determine this rank, we make use of the
rank of the reduced density matrix for the $\nu=1$ Laughlin state. We showed that to relate the
rank for the $\nu=1/m$ Laughlin state to the case $\nu=1$, one has to prove a bound on the
degree of symmetric polynomials that vanish under the formation of certain clusters. Though
we were not able to finish the proof of this statement, we made substantial progress by
explicitly constructing a set of polynomials that vanish under the clustering, and we proved
that the {\em total} degree of these polynomials is bounded from below.

We commented on a possible, though most likely rather hard, route towards finishing the proof.
In this paper, we concentrated on the Laughlin states. It would be interesting to see if similar
methods can be used to make progress on different model states, such as the Moore-Read
and Read-Rezayi states, that exhibit excitations obeying non-Abelian statistics.

\begin{acknowledgements}
The work  of B.M. and E.A. was sponsored in part by the Swedish Research Council. B.M. thanks Mohammad Khorrami and B.E.  thanks R.Santachiara, A.B.Bernevig, N.Regnault, A.Sterdyniak, P.Zinn-Justin, M.Wheeler and D.Betea for discussions.
\end{acknowledgements}

\appendix

\section{Properties of the polynomials $r_n^{(x)}$}
\label{Polynomials r_n}

In this Appendix we introduce a family of symmetric polynomials $\{ r_n^{(x)} \}$ defined through the generating function
\begin{align}
 \exp  \left ( - x \sum_{k=1}^{\infty} p_k \frac{t^k}{k} \right) = \sum_{n=0}^{\infty} r^{(x)}_n \frac{t^n}{n!}\cdot 
\end{align}
The key property of the $r_n^{(x)}$'s is their behavior under the clustering transformation $\mC_m$:
\begin{align}
\mC_m \l(r_n^{(x)}\r) = r_n^{(m\, x)},
\end{align}
which is a direct consequence of their definition. Further properties follow from the generating function, namely
\begin{itemize}
\item  $r_n^{(1)} =  (-1)^n \,n!  \, e_n$,
\item  $r_n^{(-1)} =  n!  \, h_n$ with $h_n = \sum_{\bs\lambda\vdash n} m_{\bs\lambda}$,
\item  $\left. \partial_x r_n^{(x)}\right|_{x=0} = - (n-1)! p_n$ for $n \geq 1$,
\end{itemize}
where the first two relations are obtained by comparison to the generating functions for the
$e_n$ and $h_n$, see for instance\cite{book:macdonald}.
Therefore, this family of polynomials interpolates between power sums $p_n$, elementary symmetric polynomials $e_n$, and complete homogeneous symmetric polynomials $h_n$. We give one additional property,
\begin{align}
r_n^{(x+y)} = \sum_{k=0}^n \binom{n}{k} r_k^{(x)} r_{n-k}^{(y)}, \label{x addition}
\end{align}
that follows from the definition. 

\subsection{Explicit formulas for $r_n^{(x)}$}

The generating function can be expanded using Bell's polynomials\cite{bell27}, yielding an explicit expression  for $r_n^{(x)}$, that is
\begin{align}
 r^{(x)}_n =   n ! \sum_{\bs{\lambda}\vdash n}   (-x)^{l(\bs{\lambda})} \frac{ p_{\bs{\lambda}}}{z_{\bs{\lambda}}}\cdot\label{explicit_rn}
\end{align}
Alternatively, this expression can be obtained by acting with $\mC_x$ on Newton's identity expressing elementary symmetric polynomials in terms of power sums (here, we allow $x$ to be real, and set $\mC_x p_i = x p_i$, for an infinite number of variables). Another explicit expression is given in terms of a determinant of power sums
\begin{align}
  r_n^{(x)} =   (-1)^n \left| \begin{array}{ccccc} 
  x p_1 & 1 & 0 & \cdots &  0\\  
  x p_2 & x p_1 & 2 & \ddots &  \vdots \\  
  \vdots &   & \ddots & \ddots &  0\\  
  x p_{n-1} & x p_{n-2} & \cdots & x  p_1 & n-1 \\  
   x p_n & x  p_{n-1} & \cdots &  x p_2 &  x p_1\end{array} \right|.  \label{determinant formula}
\end{align} 
The first few polynomials are given by
\begin{widetext} 
\begin{align*}
r_0^{(x)} & =1, \\
r_1^{(x)} & =-x p_1, \\
r_2^{(x)} & = x^2 p_1^2 - x p_2, \\
r_3^{(x)} & = -x ^3 p_1^3 + 3 x^2 p_1p_2 -2 x p_3,\\
r_4^{(x)} & = x^4 p_1^4 - 6x^3 p_1^2p_2 + 3x^2 p_2^2 + 8 x^2 p_1p_3 - 6x p_4, \\
r_5^{(x)} & = -x^5 p_1^5 +10 x^4 p_1^3p_2 - 15 x^3 p_1 p_2^2  - 20 x^3 p_1^2 p_3
+ 20 x^2 p_2 p_3 + 30 x^2 p_1 p_4 - 24 x p_5.
\end{align*}
\end{widetext}
From triangularity it follows that the family $\{r_1^{(x)},\ldots, r_n^{(x)} \}$ algebraically spans all symmetric polynomials in $n$ variables as long as $x\neq 0$.

\subsection{Degree of $r_n^{(x)}$}
Let us consider the monomial decomposition of $r_n^{(x)}$  
\begin{align}
r_n^{(x)} = \sum_{\bs{\lambda}\vdash n} q^{(\bs{\lambda})}_n(x) \,  m_{\bs{\lambda}}.
\label{monomial decomposition}
\end{align} 
We are going to compute the first coefficient, namely $q^{(\bs{\lambda})}_n(x) $ for $\bs{\lambda} = (n)$. If this first coefficient is non-zero, the degree of $r_n^{(x)}$ is $n$. This coefficient is a polynomial of degree $n$ in $x$, since from \eqref{explicit_rn} we have $r_n^{(x)}(z,0,0,\ldots) \sim (-x)^n z^n$ as $x$ goes to infinity.

We now take $x=m$ to be an integer, and write
$r_n^{(m)} = \mC_m \bigl( r_n^{(1)} \bigr) = (-1)^n n! \; \mC_m \bigl( e_n \bigr)$.
By considering the definition of $e_n$, 
it follows that $m \geq n$ in order for
$r_n^{(m)}(z,0,0,\ldots)$ to be non-zero, because $m_{\bs{\lambda}} = 0$ if
the number of variables is less than $l (\bs{\lambda})$. 

Therefore, $q^{(\bs{\lambda}= (n))}_n(x)$ vanishes for $x=0,1,\cdots,n-1$. It follows from the
asymptotic behavior given above that
\begin{align}
q^{(\bs{\lambda}= (n))}_n(x) = (-1)^n x (x-1) (x-2) \cdots (x-n+1).
\end{align}
Hence, the degree of $r_n^{(x)}$ is $n$ as long as $x \neq 0, 1,\cdots, n-1$. 

\subsection{Monomial decomposition of $r_n^{(x)}$}

In this Appendix we quote the full monomial decomposition of $r_n^{(x)}$, namely  
\begin{align}
q^{(\bs{\lambda})}_n(x) =  (-1)^n \binom{n}{\lambda_1,\ldots ,\lambda_n} \prod_{i=0}^{n-1} (x-i)^{\lambda_{i+1}^t}, 
\end{align} 
where $\bs{\lambda}^t$ stands for the transpose of $\bs{\lambda}$.
The parts of $\bs{\lambda}^t$ are given by
$\lambda^t_i = l(\bs{\lambda}) - \sum_{j=1}^{i-1} n_j (\bs{\lambda})$, or equivalently,
$\lambda^t_{i}-\lambda^t_{i+1} = n_i (\bs{\lambda})$.
Below, we sketch how this can be established.

\lma{
The monomial expansion of $ \left. \partial_x^j r_n^{(x)} \right|_{x=0}$ involves only partitions $\bs{\lambda}$ such that $\lambda_{j+1} \leq 0$, i.e., $\lambda_{i} =  0$ for $i \geq j+1$.
}{lma01}

\prf{
From the generating function for the $r_n^{(x)}$ we get
\begin{align}
\partial_x^j r_n^{(x)}\bigr|_{x=0} = (-1)^j n! \sum_{\substack{k_1,\ldots ,k_j \geq 1\\ k_1+\cdots+k_j=n}} \prod_i \frac{p_{k_i}}{k_i},
\end{align}
In this expression each term $\prod_i p_{k_i}$ has a monomial decomposition that involve only partitions $\bs{\lambda}$ with a length $l(\bs{\lambda}) \leq j$, from which lemma $1$ follows.}

\lma{
The monomial expansion of $ \left. \partial_x^j r_n^{(x)} \right|_{x=i}$, with $i$ an integer, involves only partitions with $\lambda_{j+1} \leq i$.}
{lma02}

\prf{
The case $i=0$ boils down to Lemma \ref{lma01}. Lemma \ref{lma02} can be proven by induction on $i$ using
\begin{align}
\partial^j_x r_n^{(x)}\bigr|_{x=i} = 
n! \sum_{p} \frac{(-1)^{n-p}}{p!}  e_{n-p} \, \l(\partial^j_x r_p^{(x)}\bigr|_{x=i-1}\r),
\end{align}
which follows from taking $\partial_x^j$ in \eqref{x addition}, namely
\begin{align}
\partial^j_x r_n^{(x+y)} = \sum_{p} \binom{n}{p}  \partial^j_x r_p^{(x)} r_{n-p}^{(y)},  
\end{align}
and then choosing $x= i-1$ and $y=1$.}

\cor{ 
The coefficient $q_n^{(\bs{\lambda})}(x)$ is of the form
\begin{align}
q_n^{(\bs{\lambda})}(x) = c_n^{(\bs{\lambda})} \prod_i  (x-i)^{\lambda_{i+1}^t}. 
\end{align}
}{cor003}

\prf{
Lemma \ref{lma02} is equivalent to stating that $\partial^j_x q^{(\bs{\lambda)}}_n (x) = 0$ for
$x=0,1,\ldots,\lambda_{j+1}-1$. Thus, $x=i$ is a root with degeneracy $\lambda^t_{i+1}$ of the polynomial in $x$ $q_n^{(\bs{\lambda})}(x)$, because $\lambda^t_{i+1}$ is the number of parts of
$\bs{\lambda}$ that are bigger or equal to $i+1$. Since this is true for all $i \in \mathbb{N}$, we have a total of $\sum_i \lambda_i^t = n$ zeros.  Since $q_n^{(\bs{\lambda})}(x)$ is of degree at most $n$, Corollary \ref{cor003} follows.}

\lma{
The coefficients $c_n^{(\bs{\lambda})}$ are given by
\begin{align}
c_n^{(\bs{\lambda})} =  (-1)^n \frac{n!}{\lambda_1! \cdots \lambda_n!} = 
(-1)^n  \binom{n}{\lambda_1,\ldots,\lambda_n}.
\end{align}
}{lma03} 

\prf{
The asymptotic behavior of \eqref{monomial decomposition} for $x$ going to infinity yields
\begin{align}
(-1)^n p_1^n = \sum_{\bs{\lambda}\vdash n} c^{(\bs{\lambda})}_n \,  m_{\bs{\lambda}}.
\end{align}
Lemma $3$ follows by expanding the left hand side using the multinomial theorem
(assuming the number of variables $p \geq n$)
\begin{align}
\left(\sum_{i=1}^{p} x_i \right)^n = \sum_{\substack{k_1, \ldots , k_p \geq 0\\k_1+\cdots +k_p = n}} \frac{n!}{k_1! \cdots k_p!} \prod_i x_i^{k_i},
\end{align}
and then gathering the terms of the r.h.s. into symmetric monomials.}

\subsection{Behavior of $r_n^{(x)}$ under translations}
\label{app:ptildederivative}

Translations are well defined in the case of finitely many variables $\{ x_1,\cdots , x_r \}$, in which case we set $p_0 =r$, i.e., the number of variables. In that case $L^{-}$ is the generator of translations
\begin{align}
L^- = \sum_{i=1}^r \frac{\partial}{\partial x_i}.
\end{align}
By Leibniz's rule, its action on $p_{\bs{\lambda}} = \prod_{j} p_{\lambda_j}$ is
\begin{align}
L^- p_{\bs{\lambda}} = \sum_{j \geq 1}  n_j(\bs{\lambda})  j p_{j-1} p_{\bs{\lambda} \setminus  \{j\}}, \qquad p_0 = r,
\end{align}
where the $n_j(\bs{\lambda})$ is the number of parts of $\bs{\lambda}$ that equal $j$, and
$\bs{\lambda} \setminus  \{j\}$ denotes the partition derived from $\bs{\lambda}$ by deleting one part that equals $j$. We can now act on $r_n^{(x)}$,
\begin{align}
L^- r_n^{(x)}  = n! \sum_{\bs{\lambda}\vdash n}
\frac{(-x)^{l(\bs{\lambda})}}{z_{\bs{\lambda}}}
\sum_{j=1}^n n_j(\bs{\lambda}) j p_{j-1} p_{\bs{\lambda} \setminus \{ j\}}.
\end{align}
We can change the summation variable from $\bs{\lambda}$ to $\bs{\mu} = \bs{\lambda} \setminus \{ j\}$, after noticing that $\bs{\mu}$ is in one-to-one mapping with $(\bs{\lambda},j)$, since
\begin{align}
j = n - |\bs{\mu}|, \qquad \bs{\lambda} = \bs{\mu} \cup \{j\}.
\end{align}
In particular $l(\bs{\lambda}) = l(\bs{\mu})+1$, $n_j(\bs{\lambda}) = n_j(\bs{\mu})+1$ and $z_{\bs{\lambda}} = z_{\bs{\mu}} (n_j(\bs{\mu})+1) j$. We find 
\begin{align}
L^- r_n^{(x)}  & = n! \sum_{\bs{\mu}\vdash 0}^{n-1} \frac{(-x)^{l(\bs{\mu})+1}}{z_{\bs{\mu}}} p_{n-1-|\bs{\mu}|} p_{\bs{\mu}}.
\end{align}
We now split the sum into two parts. First the term $|\bs{\mu}| = n-1$ is simply
\begin{align}
 n! \sum_{\bs{\mu}\vdash n-1} \frac{(-x)^{l(\bs{\mu})+1}}{z_{\bs{\mu}}} p_0 p_{\bs{\mu}}
 = -x n p_0 r_{n-1}^{(x)}.
\end{align}
Now the reminder is 
\begin{align}
 n! \sum_{\bs{\mu}\vdash0}^{n-2} \frac{(-x)^{l(\bs{\mu})+1}}{z_{\bs{\mu}}} p_{n-1-|\bs{\mu}|} p_{\bs{\mu}} = n(n-1)r_{n-1}^{(x)},
\end{align}
as can be seen from the determinant formula \eqref{determinant formula} by repeatedly developing along the last column until the matrix is 1 by 1. These last `determinants' correspond to the factor $p_{n-1-|\bs{\mu}|}$ in the sum above. Finally we get
\begin{align}
L^{-} r_n^{(x)} = n(n-1 -x p_0) r_{n-1}^{(x)},
\end{align}
from which Eq.~\eqref{eq:Lminptilde} in the main text follows.

\section{An alternate set of generators}

In this Appendix, we briefly introduce an alternate set of generators $p'_i$, that could be used in the
proof in section \ref{sec:clustering}. This set of generators of $\Lambda_{q=mN}$ is constructed
to satisfy $p'_i = p_i$ for $i \leq N$, and $\mC_m \bigl( p'_{i} \bigr) = 0$ for $i > N$, and have
total degree $i$. In the construction, to keep track of the number of variables of the power sums,
we introduce an additional index $N$, so
\begin{align}
p_{i,N}(x_1,\dots,x_N)=x_1^i+\cdots+x_N^i. 
\end{align}

Recall that $\mathcal{A}_N = \{ p_{1,N}, \cdots, p_{N,N}\}$ are algebraically independent and generates all symmetric polynomials in $N$ variables. In particular for each $i \in \mathbb{N}^*$ there exists a unique polynomial $T_{i,m,N}$
in $N$ variables, such that 
\begin{align}
m\,p_{i, N}=T_{i,m,N}(m\,p_{1,N},\dots,m\,p_{N,N}). \label{eqn:defT}
\end{align}
The $p'_{i,q}$ are now defined as follows
\eqn{
p'_{i,q}=\arr{rr}{
p_{i,q},&1\le i\le N\\\\p_{i,q}-T_{i,m,N}(p_{1,q},\dots,p_{N,q}),&N < i\le q.}}{eqn:defptilde}
By construction, they obey $\mC_m \bigl( p'_{i,q} \bigr) = 0$ for $i\geq N+1$, are non-zero,
and have total degree $i$. In addition, they form an alternate generating set 
$\mathcal{A}'_q = \{ p'_{1,q}, \ldots, p'_{q,q}\}$ of $\Lambda_{q}$, because the $p_{i,q}$ of
the generating set $\mathcal{A}_q$ can be expressed in terms of the $p'_{i,q}$ in
Eq.~\eqref{eqn:defptilde}.

Since, as is shown in Corollary \ref{cor001}, $\tilde{p}_{N+1}$ is the unique (up to a scale factor) polynomial that vanishes under
the clustering $\mC_m$, we find that $p'_{N+1}\propto \tilde{p}_{N+1}$, and that
$L^{-} p'_{N+1} = 0$. 
In addition, by a direct calculation, one finds that
$L^{-} p'_{i,q} = i p'_{i-1,q}$ for $1\leq i < N$ and $N+1 < i \leq q$.

Namely, by acting on both sides of the definition of $T_{i,m,N}$, Eq.~\eqref{eqn:defT} with
$L^-$ gives
\begin{widetext}
\begin{align*}
L^{-} T_{i,m,N}(mp_{1,N},\ldots,mp_{N,N}) &=
q T_{i,m,N;1} (mp_{1,N},\ldots,mp_{N,N}) + \sum_{j=2}^{N} 
m \,  j \, p_{j-1,N} T_{i,m,N;j} (mp_{1,N},\ldots,mp_{N,N}) \\
& = i m p_{i-1,N} = i T_{i-1,m,N} (mp_{1,N},\ldots,mp_{N,N}),
\end{align*}
\end{widetext}
where $T_{i,m,N;j}$ denotes the derivative of $T_{i,m,N}$ with respect to its $j$th argument.
In particular by setting $X_j = mp_{j,N}$,
we find that for general arguments,
\begin{widetext}
\begin{equation}
\label{eq:Tderivative}
iT_{i-1,m,N}(X_1,\ldots,X_N) =q T_{i,m,N;1}(X_1,\ldots,X_N)
+ \sum_{j=2}^{N} j X_{j-1}T_{i,m,N;j}(X_1,\ldots,X_N).
\end{equation}
\end{widetext}
We can now act with $L^{-}$ on both sides of the definition of
$p'_{i,q}$, Eq.~\eqref{eqn:defptilde}. Using the relation
Eq.~\eqref{eq:Tderivative}, we find that, for $i > N+1$,
\eqn{
L^{-} p'_{i,q}& = L^{-} p_{i,q} - L^{-} T_{i} ( p_{1,q},\ldots,p_{N,q}) \\&= 
i p_{i-1,q} - i T_{i-1} (p_{1,q},\ldots,p_{N,q})\\& = i p'_{i-1,q},
}{}
which is what we wanted to show. We note that in the case that
$i = N+1$, the only thing that changes in the argument above is that
the left hand side of Eq.~\eqref{eq:Tderivative} is replaced by
$i X_N$, leading to the result $L^{-} p'_{N+1,q} = 0$, which
we showed in the main text using a different method.

Finally, we mention that it is also possible to prove that the degree of $p'_{i,q}$ equals
$i$ for $N+1 \leq i \leq 2N+1$, directly from the definition.
We believe that the degree of $p'_{i,q}$ also equals
$i$ for $2N+1 < i \leq q$, but did not find a proof of this.

\end{document}